\documentclass[twocolumn,prl,showpacs,bibnotes]{revtex4}
\usepackage{graphicx}
\begin{document}
\draft
\def\ds{\displaystyle}
\title{Complex Spectrum of a Spontaneously Unbroken PT Symmetric Hamiltonian}
\author{C. Yuce$^\star$}
\address{Department of Physics, Anadolu University,
 Eskisehir, Turkey}
 \email{cyuce@anadolu.edu.tr}
\date{\today}
\pacs{03.65.Ge, 03.65.-w, 02.60.Lj}
\begin{abstract}
It is believed that unbroken PT symmetry is sufficient to
guarantee that the spectrum of a non-Hermitian Hamiltonian is
real. We prove that this is not true. We study a Hamiltonian with
complex spectrum for which PT symmetry is not spontaneously
broken.
\end{abstract}
\maketitle

\section{Introduction}

It is one of the postulates of quantum mechanics that any operator
associated with a physically measurable property is Hermitian.
Hermiticity is a useful property as it guarantees the reality of
the spectrum. However, it was shown in 1998 that the following
non-Hermitian potential
\begin{equation} \label{benpot}
U(x)=x^2 (ix)^{\epsilon}~~~~~~~~~(\epsilon \geq 0)~,
\end{equation}
have real positive spectra \cite{b1}. It was realized that
although this potential is not Hermitian, it is invariant under
$PT$ transformations, where $P$ and $T$ are linear parity and
antilinear time-reversal operator, respectively. They have the
effect
\begin{equation}\label{PTcondition}
p\rightarrow p,~~~~x\rightarrow -x,~~~~i\rightarrow
-i,~~~~t\rightarrow -t.
\end{equation}
Hermiticity requirement is replaced by the analogous condition of
space-time reflection symmetry. It should be noted that $PT$
symmetry can not be regarded as the fundamental property as there
exist also examples with real spectra for which not even the
Hamiltonian is $PT$ symmetric. Furthermore, there are also
PT-symmetric Hamiltonians that do not have a real spectrum.
Neither Hermiticity nor $PT$ symmetry serves as a sufficient
condition for a quantum Hamiltonian to
preserve the reality of energy spectrum.\\
For the case of a PT symmetric Hamiltonian, the PT operator
commutes with the Hamiltonian $H$. Because PT operator is not
linear, the eigenstates of H may or may not be eigenstates of PT .
If all eigenfunction of a PT symmetric Hamiltonian is also an
eigenfunction of the PT operator, it is said that PT symmetry of H
is unbroken.
\begin{equation}\label{ptcond12}
[H,PT]=0~;~~~~~ PT \Psi(x)=\mp ~\Psi(x)~.
\end{equation}
Conversely, if some of the eigenfunctions of a PT symmetric
Hamiltonian are not simultaneously eigenfunctions of the PT
operator, then the PT symmetry of H is spontaneously
broken.\\
The appearances of complex eigenvalues in the spectra of
PT symmetric quantum mechanical systems are usually associated
with a spontaneous breaking of PT transformations. In other words,
the eigenvalues are real when the Hamiltonian and its wave
function remain invariant under a simultaneous parity and time
reversal transformations \cite{b1,spontan1,spontan2,spontan3,japar}.\\
As an example, the potential (\ref{benpot}) is $PT$ symmetric
because it is invariant under (\ref{PTcondition}). The energy
spectrum is not real when $\ds{\epsilon < 0}$. This is because
$\ds{\epsilon < 0}$ is the parametric region of
broken PT symmetry \cite{b1}.\\
In this study, in contrast with the first conjectures, we will
show that unbroken $PT$ symmetry is not sufficient to guarantee
that the spectrum of a non-Hermitian
Hamiltonian is real.\\
This paper is organized as follows. Section 2 considers a
non-Hermitian Hamiltonian and finds its eigenstates and complex
eigenvalues exactly. Section 3 discusses the reason for the
existence of complex energy spectrum when  $PT$ symmetry is not
spontaneously broken. Finally, section 4 concludes our discussion.

\section{Formalism}

Consider the following non-Hermitian Hamiltonian
\begin{equation}\label{e1}
H=p^2+  x^2+2i f(t)~ x~,
\end{equation}
where $\ds{f(t)}$ is a real-valued time-dependent function. The
constants are set to unity for simplicity $(\hbar^2/2m=\hbar=1)$.
The Hamiltonian is invariant under $PT$ transformations provided
that $\ds{f(-t)=f(t)}$. This analytically solvable Hamiltonian
allows us to investigate the relation between the reality of
the energy spectrum and the broken $PT$ symmetry.\\
Suppose first that $\ds{f(t)}$ is equal to a constant $\ds{f_0}$.
In this case, the eigenfunctions of the Hamiltonian can be readily
found by shifting the coordinate as $\ds{x\rightarrow
x^{\prime}=x+if_0}$. The potential in the primed coordinate system
reads $\ds{U(x^{\prime})={x^{\prime}}^2+f_0^2}$. The boundary
conditions remain unchanged and the energy eigenvalues are shifted
by a constant $\ds{f_0^2}$ and given by $\ds{E_n=2n+1+f_0^2}$,
where $n=0,1,...$. As can be seen, the energy eigenvalues are real
when $\ds{f(t)=f_0}$. As a result, it can be claimed that since
$PT$ symmetry is not spontaneously broken for this case, the
energy eigenvalues are real.\\
Suppose next that $f(t)$ depends on time explicitly. Let us now
obtain the exact analytic solution and investigate whether the
reality of energy spectrum is maintained. The corresponding
Schrodinger equation reads
\begin{equation}\label{deriv}
-\frac{\partial^2 \Psi}{\partial x^2}+  \left(x^2+2i f(t) x\right)
\Psi =i \frac{\partial \Psi}{\partial t} ~,
\end{equation}
The following coordinate transformation is introduced to solve
this equation
\begin{equation}\label{cgh}
z=x+ig(t),
\end{equation}
where the real-valued time-dependent function $g(t)$ to be
determined later. Under the coordinate transformation (\ref{cgh}),
the time derivative operator transforms as $\ds{\partial/\partial
t \rightarrow
\partial/\partial t+i\dot{g}~
\partial/\partial z}$, where dot denotes time derivation. Substitution of the equation (\ref{cgh}) into the equation
(\ref{deriv}) yields
\begin{equation}\label{dsereriv2}
-\frac{\partial^2 \Psi}{\partial z^2}+\dot{g} \frac{\partial
\Psi}{\partial z} + \left(z^2+2i (f-g)z+ (2f-g)g\right) \Psi =i
\frac{\partial \Psi}{\partial t} ~.
\end{equation}
Let us introduce
\begin{equation}\label{paramfr}
\Psi(z,t)=e^{\alpha(t) ~z }~~ \Phi(z,t)~,
\end{equation}
where  $\alpha(t)$ is a time-dependent function. The equation
(\ref{dsereriv2}) simplifies if we choose $\ds{\alpha(t)}$ and
$\ds{g(t)}$ as
\begin{equation}\label{cond}
\alpha=\frac{\dot{g}}{2}~;~~~~~~~~~ \dot{\alpha}=2(f-g) ~.
\end{equation}
Substituting (\ref{paramfr},\ref{cond}) into (\ref{dsereriv2}), we
obtain
\begin{equation}\label{dserer3}
-\frac{\partial^2 \Phi}{\partial z^2}+
\left(z^2+(2f-g)g+\frac{\dot{g}^2}{4}\right) \Phi =i
\frac{\partial \Phi}{\partial t} ~.
\end{equation}
Separation of variable techniques can be applied to get rid of the
time-dependent part.
\begin{equation}\label{timeridof}
\Phi(z,t)=\exp{\left(-iE_nt-i\int_0^t
\left((2f-g)g+\frac{\dot{g}^2}{4}\right) dt^{\prime}\right)}~~
\phi(z)~,
\end{equation}
Then, the equation (\ref{dserer3}) is reduced to
\begin{equation}\label{dssdas3}
-\frac{d^2 \phi}{dz^2}+ z^2 \phi =E_n\phi~,
\end{equation}
It has been shown that the time dependent part of the Schrodinger
equation (\ref{deriv}) can be eliminated and the problem is
reduced to the well-known harmonic oscillator one. So, the
constant $E_n$ is given by
\begin{equation}\label{dfgc3}
E_n=2(n+\frac{1}{2})~,
\end{equation}
where $n=0,1,2,...$.\\
Using the eigenfunctions for the standard harmonic oscillator
(\ref{dssdas3}) and transforming backwards yields the exact
eigenfunctions for the Hamiltonian (\ref{e1})
\begin{eqnarray}\label{sonuc}
\Psi_n=\exp{\left(-iE_nt-i\int_0^t
\left((2f-g)g+\frac{\dot{g}^2}{4}\right) dt^{\prime}\right)}
\times \nonumber \\
\exp{\left(\alpha (x+ig)-(x+ig)^2/2\right)} ~H_n{(x+ig)},
\end{eqnarray}
where $H_n(x+ig)$ are the Hermite polynomials. The first two of
them are given by $\ds{H_0(x+ig)=1}$ and $\ds{H_1(x+ig)=2(x+ig)}$. \\
Having obtained the exact analytic solution, let us find the
energy eigenvalues for the time-dependent non-Hermitian
Hamiltonian (\ref{e1}). \\
The ground state energy is computed as
\begin{equation}\label{sdrl}
<0|E|0>=\frac{\int_{-\infty}^{\infty}  \Psi_0^{\star}~H\Psi_0
~dx}{\int_{-\infty}^{\infty} |\Psi_0|^2
dx}=1+g^2+\alpha^2+2i\alpha f~.
\end{equation}
and the energy eigenvalue for the first excited state is given by
\begin{equation}\label{sdDcx}
<1|E|1>=\frac{E_{11}+2i\alpha f
(3+2\alpha^2+2g^2)}{1+2g^2+2\alpha^2}~,
\end{equation}
where $\ds{E_{11}=3+7(g^2+\alpha^2)+4\alpha^2
g^2+2(\alpha^4+g^4)}$ is the real part of the numerator. The
energy eigenvalues for the excited states can also be found
exactly.\\
As it can be seen from  (\ref{sdrl},\ref{sdDcx}), the energy
eigenvalues for the ground and the first excited states are real
if and only if either $f(t)$ or $\alpha(t)$ is equal to zero. The
former one is trivial since the Hamiltonian (\ref{e1}) becomes a
Hermitian operator when $f(t)$ is zero. The latter case is
surprising. The time-dependent function $\alpha(t)$ is equal to
zero if $f(t)$ is a constant ( Suppose $f(t)=f_0$, then using
(\ref{cond}) we get $g(t)=f_0$ and $\alpha(t)=0$). So, the energy
spectrum for the non-Hermitian Hamiltonian is not real unless
$f(t)=0$ or $f(t)=f_0$.\\
If $\alpha(t)$ does not vanish ($\ds{f(t)}$ depends on time
explicitly) then the corresponding energy eigenvalues are not real
any more. As an example, let the non-Hermitian potential be given
by $\ds{V(x)=x^2+2it x}$. In this case, $\ds{f(t)=t}$ and using
(\ref{cond}) we get $\ds{g(t)=t}$ and $\ds{\alpha=1/2}$.
Substitution these into (\ref{sonuc}), we get the exact wave
function
\begin{equation}\label{speccase}
\Psi_n=e^{\left(-2i(n+\frac{1}{2})t-i
\left(\frac{t^3}{3}+\frac{t}{4}\right)+(x+it)/2-(x+it)^2/2
\right)}~ ~H_n{(x+it)}
\end{equation}
The energy eigenvalues for the ground and the first excited states
are given by
\begin{equation}\label{fgch}
<0|E|0>=t^2+\frac{5}{4}+it~.
\end{equation}
\begin{equation}\label{fgch1}
<1|E|1>=\frac{39+16(4t^2+t^4)+4i (7t+4t^3) }{4 (3+4t^2)}~.
\end{equation}
As a second example, let us choose $\ds{f(t)=t^2}$. If we solve
(\ref{cond}), we obtain $\ds{g(t)=t^2-1/2}$ and $\ds{\alpha=t}$.
The exact wave function for the non-Hermitian potential
$\ds{V(x)=x^2+2it^2 x}$ reads
\begin{eqnarray}\label{speccase2}
\Psi_n(x,t)=\exp{\left(t
(x+i(t^2-\frac{1}{2}))-\frac{1}{2}(x+i(t^2-\frac{1}{2}))^2\right)}\nonumber\\
\times~e^{-2i(n+\frac{1}{2})t-i~
(\frac{t^5}{5}+\frac{t^3}{3}-\frac{t}{4})}~~H_n{\left(x+i(t^2-\frac{1}{2})\right)}.~~~~~~~~~~
\end{eqnarray}
This example is of great importance since $PT$ symmetry is not
spontaneously broken for this case. It was conjectured that the
reality of the spectrum was due to the unbroken $PT$ symmetry of
the non-Hermitian Hamiltonians. So, one expects that the energy
spectrum is real since the Hamiltonian and its eigenfunction
(\ref{speccase2}) are both invariant under $PT$ operations.
However, the energy eigenvalue for the ground state is not real as
can be seen below
\begin{equation}\label{fsgfh}
<0|E|0>=t^4+\frac{5}{4}+2it^3~.
\end{equation}
One can run Mathematica program to calculate the energy
eigenvalues for the excited states. They are also not real. The
general idea that the energy eigenvalues are real for unbroken
$PT$ symmetric Hamiltonian does not hold for this example.\\
In the next section, we will explain why complex eigenvalues
appear for the non-Hermitian Hamiltonian for which $PT$ symmetry
is not spontaneously broken.

\section{Symmetry}

As has been shown, the condition (\ref{ptcond12}) to guarantee the
reality of energy spectrum does not always work.  So, we need a
more general condition to explain the reason for the existence of
complex energy eigenvalues when PT symmetry of a Hamiltonian H is
not spontaneously broken. \\
Let a Hamiltonian be given  by $\ds{H
=\frac{\mathbf{p}^2}{2m}+U^{R}+i U^{I}}$, where the real-valued
functions $U^{R}(x,t)$, $U^{I}(x,t)$ are the real and the
imaginary parts of the Hamiltonian H, respectively. It was
suggested in \cite{cem} that the energy spectrum for this
non-Hermitian Hamiltonian is real if the expectation value of its
imaginary part, $\ds{<U^{I}>}$, is equal to zero
\begin{equation}\label{asisrtgl}
< U^{I}>=\int |\Psi | ^2~  U^{I}~d^3 x=0~,
\end{equation}
where the integral is taken over all space.\\
Hermiticity and $PT$ symmetry conditions can be derived from
(\ref{asisrtgl}) as special cases. Trivially, when a Hamiltonian
is Hermitian,
$\ds{U^{I}=0}$, then (\ref{asisrtgl}) is satisfied. The latter one will be derived in the next section. \\
Let us calculate $\ds{<U^{I}>}$ for the Hamiltonian (\ref{e1}). It
is given by
\begin{equation}\label{45sdhgx}
<2 f(t) ~x>=2 f(t)\int_{-\infty}^{\infty} \Psi^{\star} ~x~ \Psi
~dx ~.
\end{equation}
If we use the exact wave function (\ref{sonuc}), we get
\begin{equation}\label{4vcngx}
2 f~e^{g^2}\int_{-\infty}^{\infty} e^{-x^2+2\alpha x}~~x~H_n(x-ig)
~H_n(x+ig) ~dx ~.
\end{equation}
Since the product of the Hermite polynomials $\ds{H_n(x-ig)
~H_n(x+ig)}$ is even for all $\ds{n}$, the above integration is
equal to zero when $\ds{\alpha=0}$. This is because the integral
of an odd function on a symmetric interval is zero.
\begin{equation}\label{465iox}
<U^{I}>=2 f(t) <x>=0~, ~~~~~~~~if~~ \alpha=0
\end{equation}
Note that $\alpha=0$ when $f(t)=0$ or $f(t)=f_0$. It is concluded
that the energy spectrum is real if $f(t)$ does not depend on time
explicitly since $\ds{<U^{I}>}$ vanishes for these cases. When
$\ds{\alpha \neq 0}$ the odd character of the integrand is lost
and the integral is not in general equal to zero. Hence, the
energy spectrum is not real any more. \\
The theorem given in \cite{cem} successfully explains the reason
of getting complex energy eigenvalues when $f(t)=t$ and
$\ds{f(t)=t^2}$. $PT$ symmetry condition is not successful to
explain the complex energy eigenvalues for $f(t)=t^2$.

\section{Conclusion}

A question arises. Is there a condition that guarantees that
$\ds{<U^{I}>}$ vanishes for a given Hamiltonian. We will conclude
our study by showing the existence of such a condition.\\
Let us replace the $PT$ symmetry condition (\ref{ptcond12}) by the
following ones.
\begin{eqnarray}\label{heg1}
U^{I} (-x,t)~~&=&-U^{I} (x,t)~;\nonumber \\
\mid \Psi(-x,t) \mid ^2 &=& \mid \Psi(x,t) \mid ^2~.~
\end{eqnarray}
Let us now find the expectation value $\ds{<U^{I}>}$ for this
case. It is given by
\begin{equation}\label{heg3}
< U^{I}>=\int_{-\infty}^{\infty}  \mid \Psi(x,t) \mid
^{2}~U^{I}(x,t)~dx ~.
\end{equation}
Let us change the variable $x$ in the integral (\ref{heg3}) to
$\ds{-x}$. It can be seen that the function in the integral is an
odd function of $x$. Since the integral of an odd function on a
symmetric interval is zero, $\ds{<U^{I}>=0}$. So, it is concluded
that if (\ref{heg1}) is satisfied than the corresponding energy
eigenvalues are real for the non-Hermitian Hamiltonians.\\
All spontaneously unbroken $PT$ symmetric Hamiltonians satisfy the
condition (\ref{heg1}) if the potential is time-independent. This
is why $PT$ symmetry condition works for the time-independent
non-Hermitian Hamiltonians.\\
On the contrary, they make difference for the Hamiltonians which
explicitly depend on time \cite{dutra,cemyuce}. Even if the
Hamiltonian and the corresponding wave function $\Psi(x,t)$ are
invariant under $PT$ transformations, they don't necessarily
satisfy the above condition (\ref{heg1}). For example, the
potential $\ds{V(x)=x^2+2it^2 x}$ and the exact wave function
(\ref{speccase2}) are both invariant under $PT$ transformations,
but they fail to satisfy the above condition (\ref{heg1}). So, the energy spectrum is not real for that case.\\
To sum up, $PT$ symmetry requirement coincides with the condition
(\ref{heg1}) for the time-independent potential fields. So far, we
have considered that the interval (boundaries) is symmetrical. If
it is not so, the energy spectrum may not be real even the
equation (\ref{heg1}) is satisfied. So, for the investigation of
the reality of energy eigenvalues for a non-Hermitian Hamiltonian,
it is better to use the condition (\ref{asisrtgl}).

\end{document}